\documentclass[11pt,twoside]{article}

\usepackage{asp2006}
\usepackage{epsf}
\usepackage{psfig}
\usepackage{lscape}
\usepackage{graphicx}

\begin{document}
\title{The Continuing Saga of the Explosive Event(s) in the M87 Jet:
  Is M87 a Blazar?}

\author{D. E. Harris} 

\affil{SAO, 60 Garden St. Cambridge, MA 02138}

\author{C. C. Cheung\altaffilmark{1}, \L. Stawarz}

\affil{KIPAC, Stanford University, Stanford CA 94305} 

\author{J. A. Biretta and W. Sparks} 

\affil{STScI, 3700 San Martin Drive, Baltimore, MD 21218}

\author{E. S. Perlman} 

\affil{Florida Institute of Technology, 150 West University Boulevard,
Melbourne, FL 32901} 

\author{A. S. Wilson} 

\affil{Astronomy Dept., University of Maryland, College Park, MD 20742} 

\altaffiltext{1}{Jansky Postdoctoral Fellow of the National Radio
Astronomy Observatory}

\begin{abstract} 
We review the recent data on the knot HST-1 in the M87 jet in the
context of typical blazar behavior.  In particular we discuss the
wide-band flare of 2005 which peaked at a factor of 50 to 80 times the
intensity observed in 2000; the superluminal radio features; and the
arguments that support the hypothesis that HST-1 was the source of the
excess TeV emission found by H.E.S.S. in 2005.  To the extent that M87
can be classified as a blazar, perhaps observed at a somewhat larger
angle to the line of sight compared to most blazars, all of these
blazar properties originate at a distance greater than 100 pc from the
nucleus, and thus cannot be associated with the location of the
'launching of the jet'.
\end{abstract}


\section{Introduction}   
\subsection{Preamble}   

\begin{figure}[!ht]
\centerline{\includegraphics[scale=0.4]{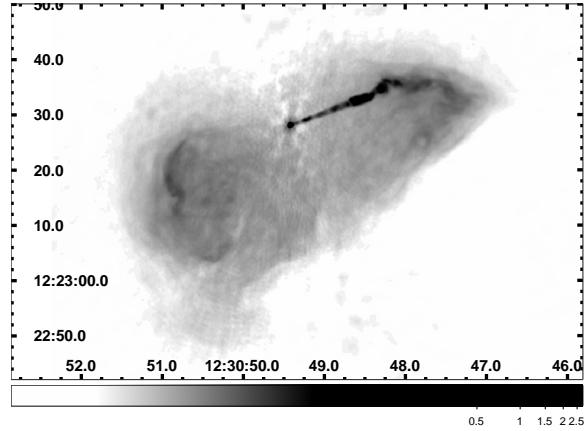}}
\caption{An 8 GHz image from the VLA of the inner part of M87 (Owen,
  private communication).}\label{fig:xband}
\end{figure}

As a reminder: when viewing a radio image (fig.~\ref{fig:xband}) of a
non-thermal source, we think we see a (distorted) view of an emitting
volume containing relativistic electrons and magnetic fields.

\begin{itemize}

\item{the morphology is distorted by projection effects \& relativistic aberration}

\item{the intensity is distorted by relativistic beaming}

\end{itemize}

\begin{figure}[!ht]
\plottwo{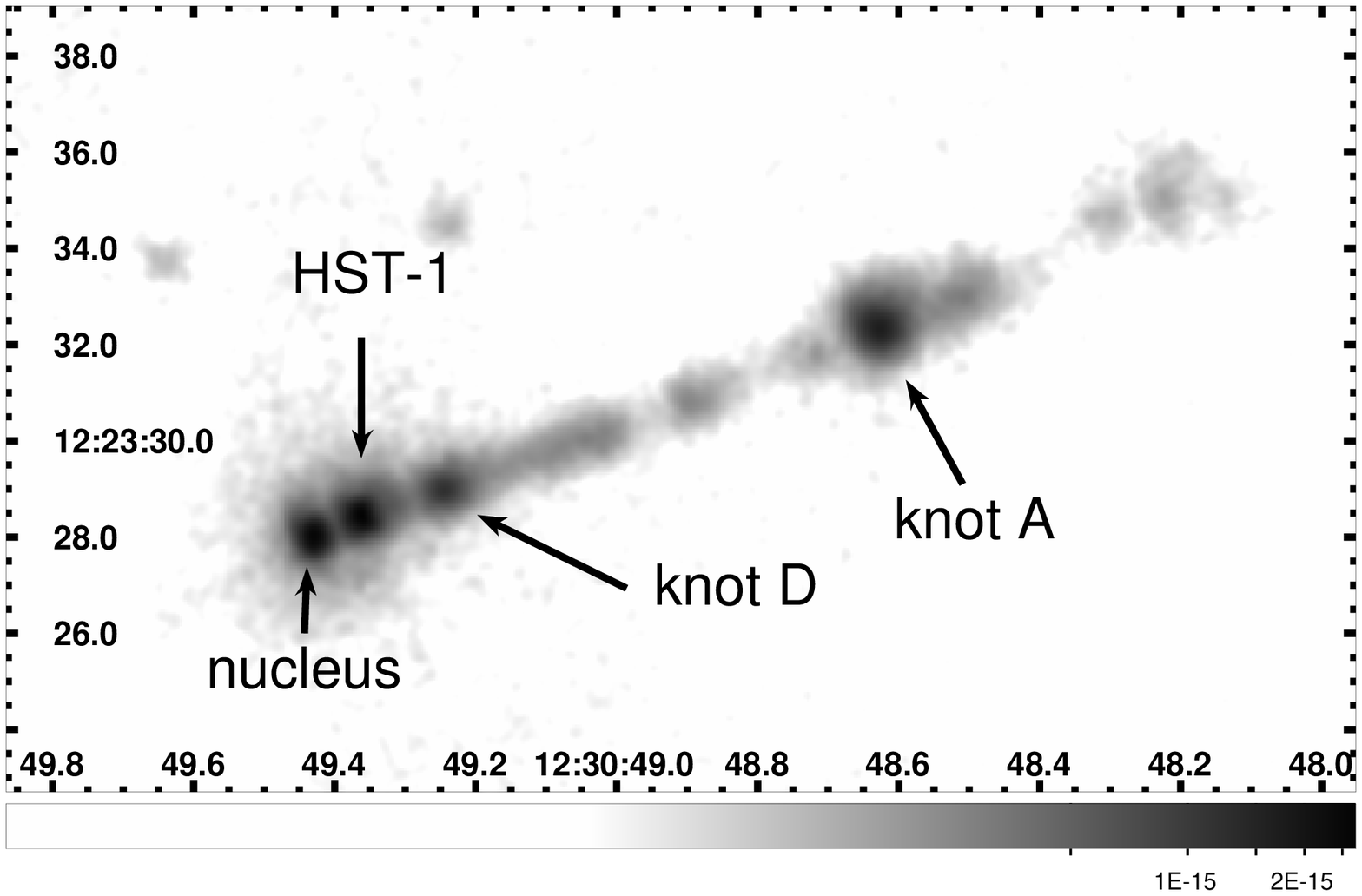}{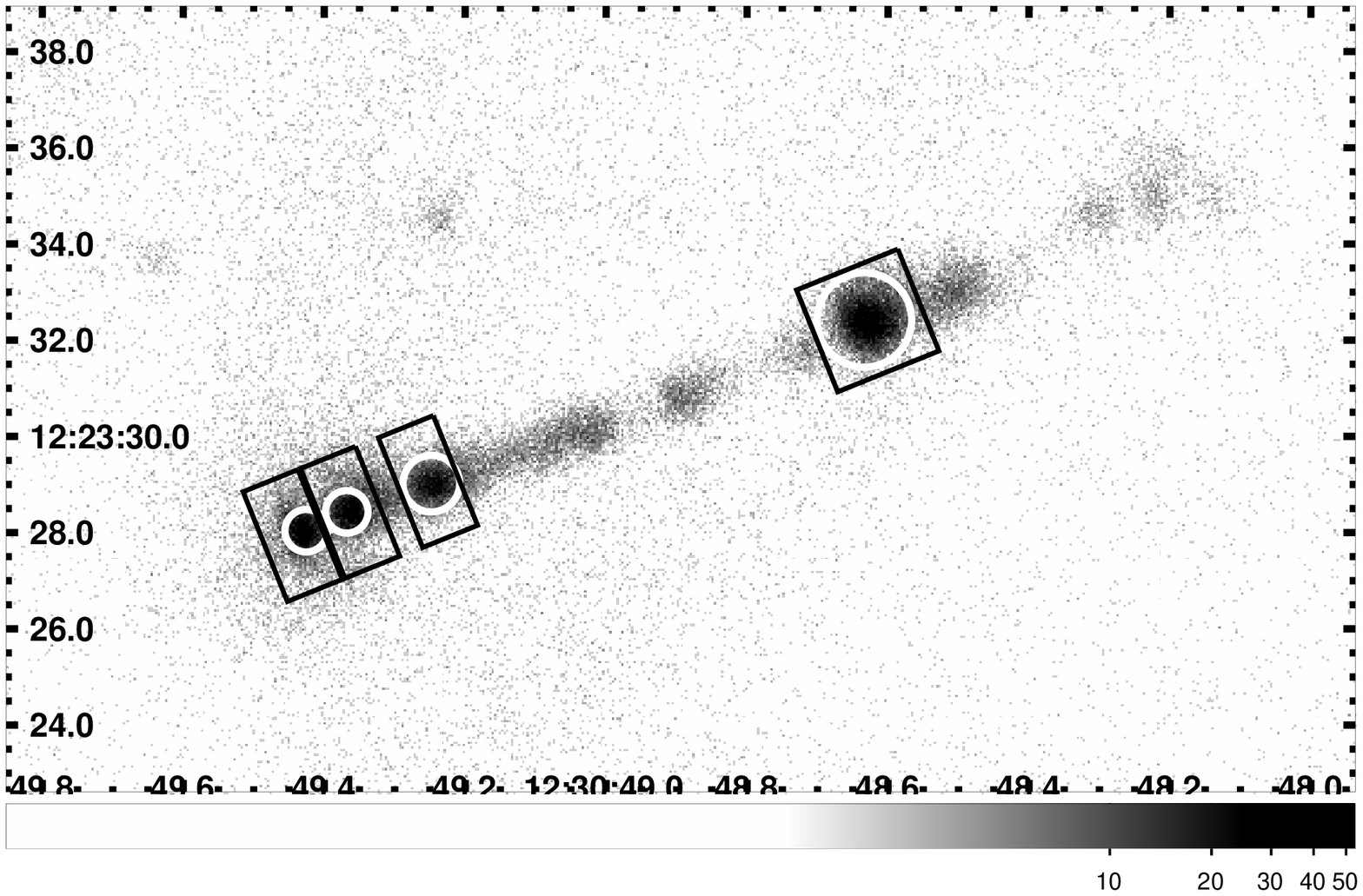}
\caption{[Left] An X-ray 'fluxmap' of the M87 jet taken from a long archival
  observation.  The energy band covered is 0.2 to 6 keV.  Pixel
  randomization has been removed and a Gaussian of
  0.25$^{\prime\prime}$ has been applied.  [Right] The event file for
  the same observation.  Two sets of regions are shown for the 4
  features measured:  small circles  and larger rectangles designed to include a
  larger fraction of the PSF.}\label{fig:chanlong}
\end{figure}

However, when viewing an X-ray image (fig.~\ref{fig:chanlong}) of an
FRI jet which is believed to also represent synchrotron emission
(Harris \& Krawczynski 2006), in addition to the same attributes as the
radio, all emission regions are also acceleration regions: i.e. high
energy electrons are being produced throughout the emitting volumes.

\subsection{History}

The initial description of X-ray variability in the M87 jet was published in
Harris et al. (2003).  This was followed by Perlman et al. (2003) on
the corresponding optical variability.  The major flare of HST-1 was
described in Harris et al. (2006) and the VLBA results on HST-1 were
published in Cheung, Harris, and Stawarz (2007).

\subsection{This paper}

In this contribution we review the evidence that M87 resembles a
blazar.  The basic argument hinges on 3 attributes: (1) the flaring
behavior of the knot HST-1; (2) the superluminal proper motions found in
HST-1; and (3) the probable association of enhanced TeV emission with
HST-1.  If M87 were to be much further away than it is, so that HST-1
would not be resolved from the nucleus, all of these phenomena would
be interpreted as coming from the immediate vicinity of the super
massive black hole (SMBH) rather than from a feature in the jet lying
more than 100 pc from the SMBH.  Insofar as we have found blazar
behavior arising from a knot in the jet, it seems clear that the basic
blazar paradigm needs substantial revision.

We first describe the X-ray lightcurves of the nucleus, HST-1, knot D
\& knot A; then the VLBA data on superluminal motions of components
within HST-1; and finally, the TeV connection.  We take the distance
to M87 to be 16 Mpc so that 1$^{\prime\prime}$=77pc.

\section{X-ray Photometry and Light Curves} 

It is relatively straight forward to separate the core from HST-1 when
there is no pileup (fig.~\ref{fig:chanlong}) since we can remove pixel
randomization and model the Point Spread Function (PSF).  To first
order we can also recover the energy of the piled events by weighting
each event with its energy and summing 0.2-17 keV.  We can also
recover the events rejected in standard processing because of grade
migration by using the evt1 file with no grade filtering.

\begin{figure}[!ht]
\plottwo{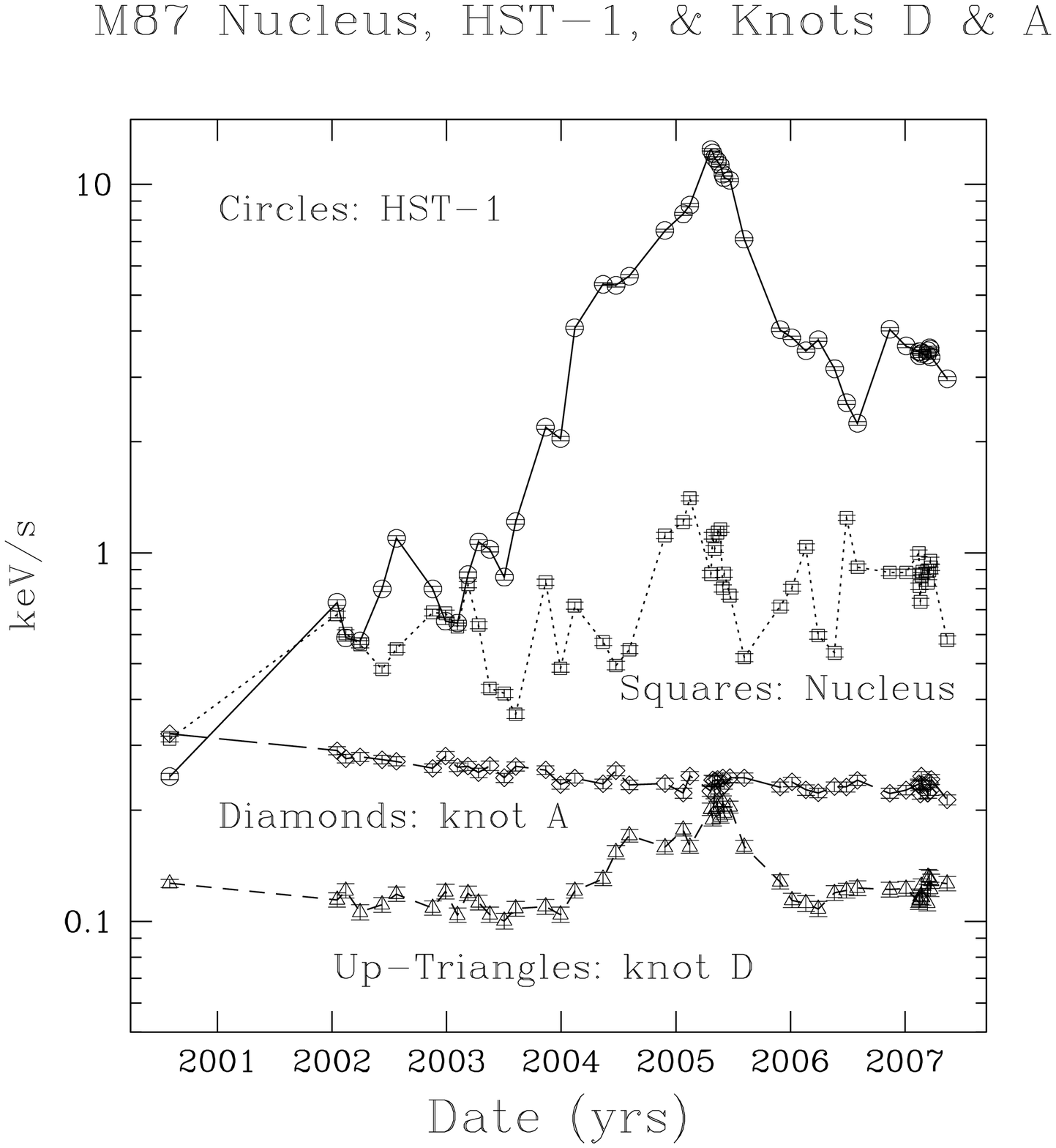}{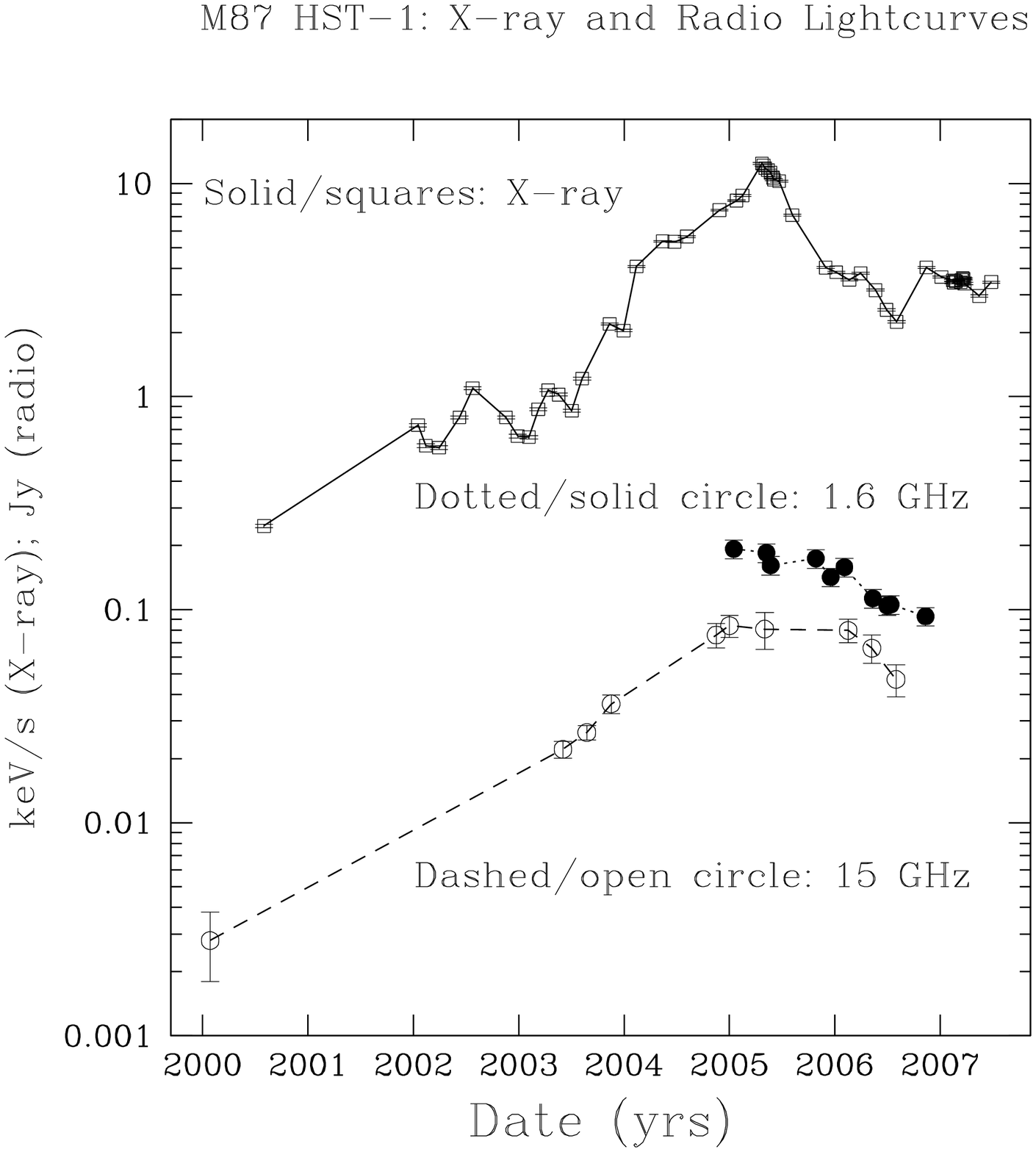}
\caption{[Left] The X-ray lightcurves on a log scale for the nucleus, HST-1,
  knots D \& A.  [Right] The radio and X-ray lightcurves of HST-1}\label{fig:lc}
\end{figure}

The resulting lightcurves are shown in fig.~\ref{fig:lc} (left panel).
In April 2005, HST-1 attained a peak intensity more than 50 times that
observed in 2000 July.  The nucleus is variable but does not display
any sort of similar event.  Knot A shows a slow decline, most likely
caused by the buildup of contaminant of the ACIS detector: the
'light-bucket' method of measuring the total charge (ev/s) does not
incorporate the changing effective area (see Harris et al. 2006).

\begin{figure}[!ht]
\plottwo{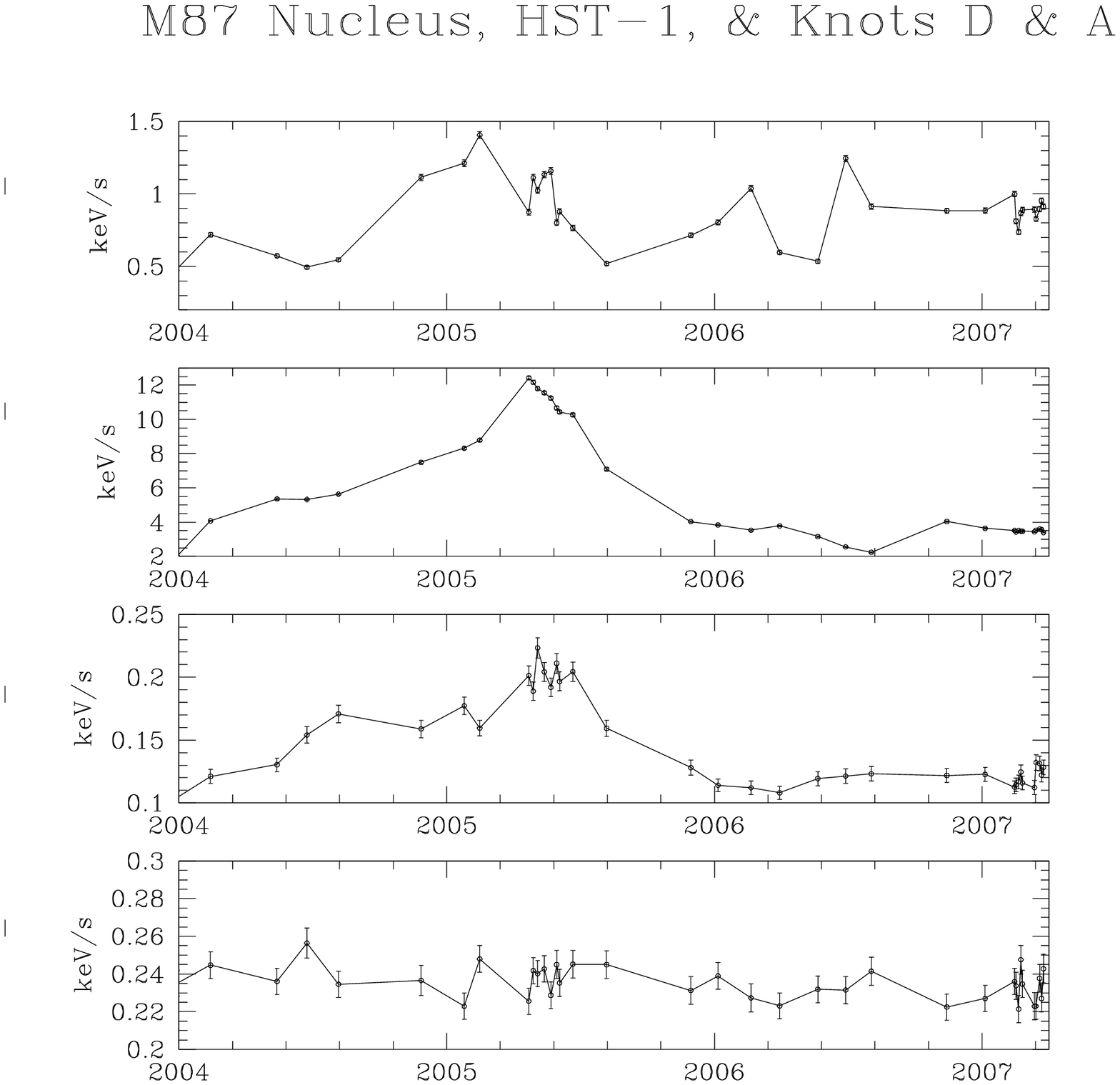}{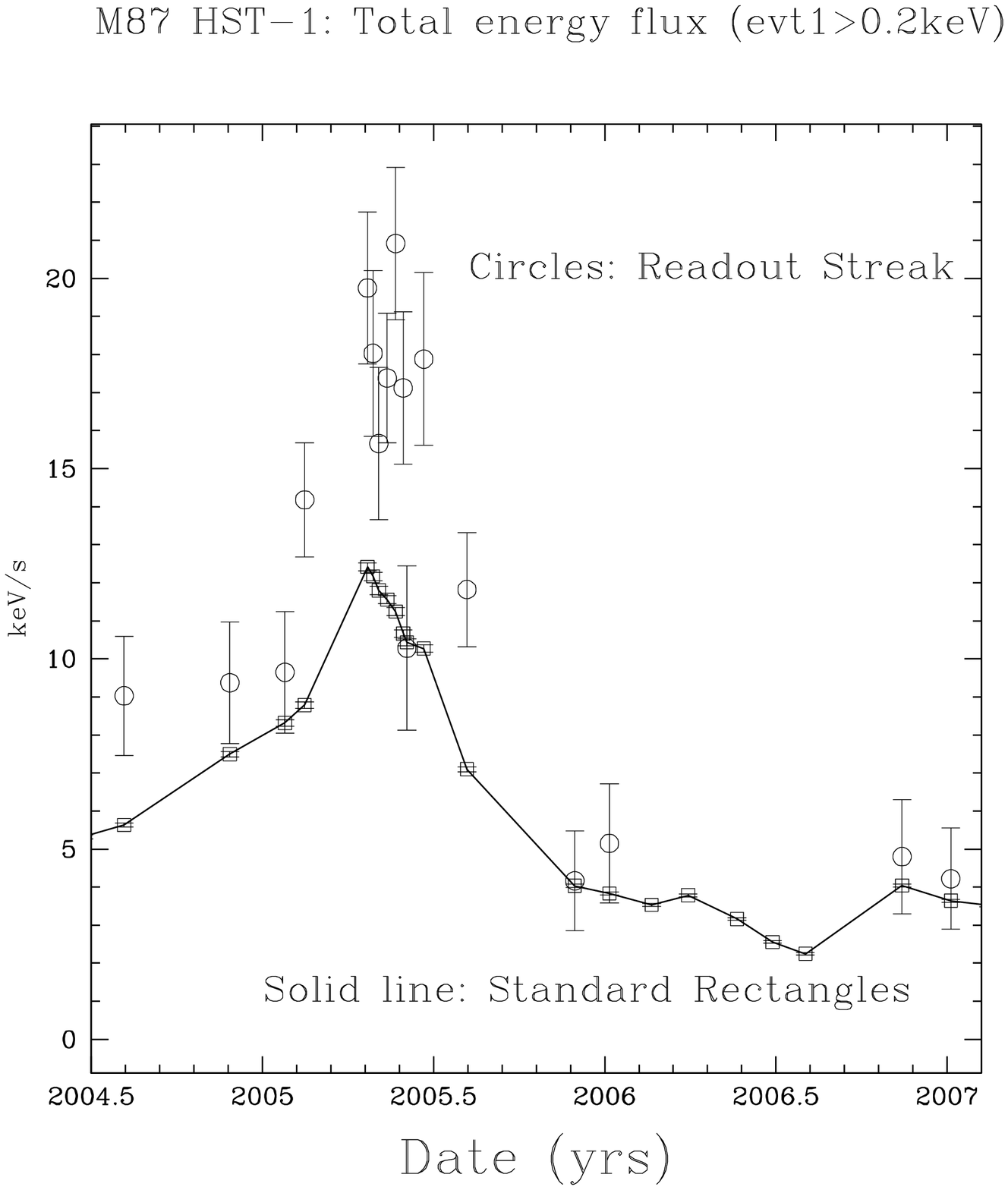}
\caption{[Left] The light curves for (top to bottom) the nucleus,
  HST-1, knot D, and knot A.  The peak in the top panel early in 2005
  and that in knot D in mid 2005 are likely caused by a combination of
  the wings of the PSF of HST-1 and by contamination via 'release of
  trapped charge' from HST-1.  [Right] The lightcurve for HST-1 with
  photometry of the readout streak included for those epochs where the
  streak was measurable.  The uncertainties are much larger than those
  of the standard measurements because of the short effective exposure
  time ($\approx$ 18s for a 5ksec observation) and the higher
  background in the larger photometric aperture.}\label{fig:4pan}
\end{figure}

The major flare of HST-1 and its affect on the nucleus and knot D are
shown in fig.~\ref{fig:4pan} (left panel).  Although the major hump in
the light curve of the nucleus in early 2005 can be mostly smoothed
out by subtracting 5\% of HST-1, another effect, the 'bleeding' from
release of trapped charge at readout time causes a secondary
enhancement to the side of the PSF away from the readout buffer.
Since this asymmetric effect changes location on the sky with the roll
angle of the satellite, during an M87 observing year (Nov-Aug) it
contaminates first the nucleus and then knot D.  The end of March is
when the readout direction is perpendicular to the jet so that is the
separation date and explains the offsets between the contaminated
lightcurves of the nucleus and knot D compared to HST-1.

Other losses which are difficult to correct include 'eat-thy-neighbor'
(a photon from the nucleus falls within the 3x3 pixel region of one
from HST-1 during the same frame time), on-board rejection of some
grades, and when part of the charge of piled events falls outside the
3x3 pixel region.

The only recourse to estimate these losses is readout streak
photometry.  Although this entails poor s/n since the effective
background is high and the effective exposure is short, in
fig.~\ref{fig:4pan} (right panel) we show that the true peak of HST-1
is significantly higher than what we had thought.

\section{Superluminal proper motions in HST-1}

We started monitoring the M87 jet with the VLA\footnote{The National
  Radio Astronomy Observatory is a facility of the National
Science Foundation operated under cooperative agreement by Associated
Universities, Inc.} in 2003 at 8 GHz and
higher in the A and B configurations which provide sufficient angular
resolution to separate HST-1 from the nucleus.  To avoid data gaps
when the VLA is in C \& D arrays, we started VLBA monitoring in 2005,
primarily at 1.6 GHz.  The resulting lightcurves for HST-1 are shown
in fig.~\ref{fig:lc} (right panel), where it can be seen that the X-ray flare was
roughly duplicated in the radio band (the optical lightcurve is
similar).

\begin{figure}[!ht]  
\centerline{\includegraphics[scale=0.6]{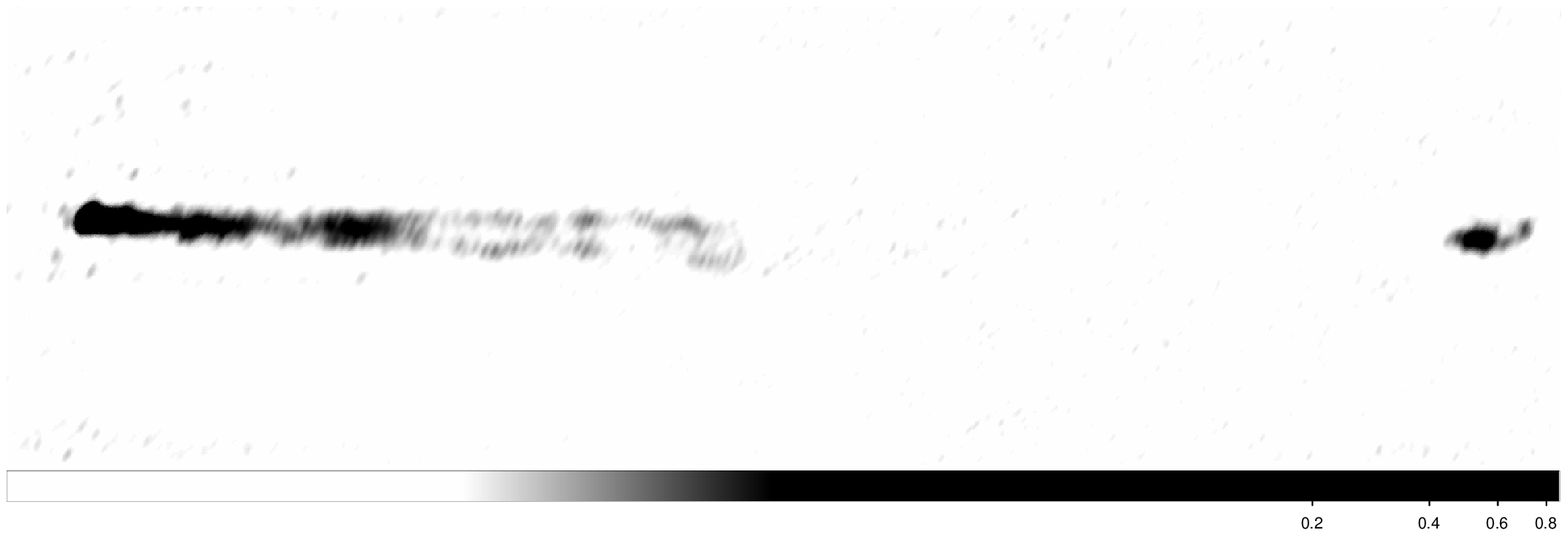}}
\centerline{\includegraphics[scale=0.6]{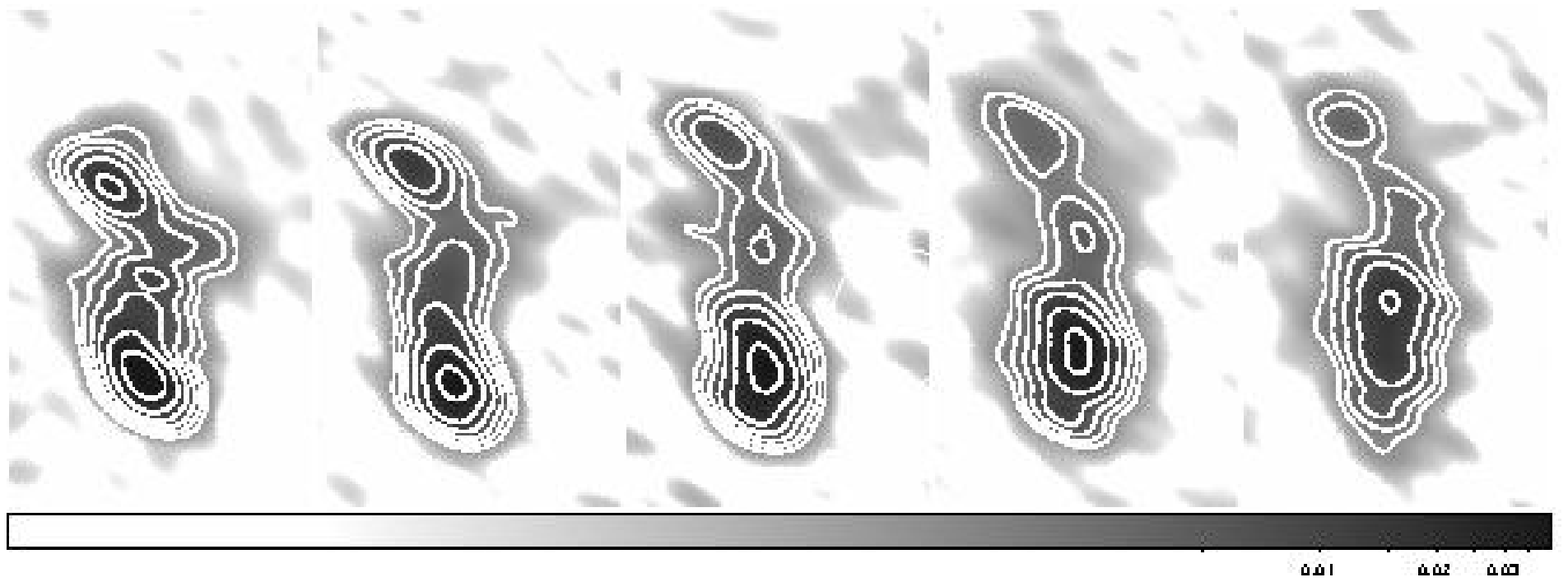}}
\caption{[Top] The VLBA image of the M87 jet out to HST-1
  ($\approx0.84^{\prime\prime}$ from the nucleus).  The observation
  is at 1.6 GHz with a clean beam of 8x3.4 mas (PA=0).  The jet has
  been rotated by 25$^{\circ}$ and was observed in 2006 June.
  [Bottom] 5 epochs of HST-1 from 2005 Jan to 2006 Jun.  Contours
  start at 1 mJy/beam and increase by factors of two.  The beamsize is
  the same as the upper panel.  The maps have been rotated by
  65$^{\circ}$ so that the bottom of each image is the upstream edge.}\label{fig:vlba}
\end{figure}

The inner (VLBA) jet is not known to be superluminal and we did not
detect any obvious motion although the jet has a quasi-continuous
structure and it is difficult to identify distinct components.
Instead of an unresolved structure for HST-1, the VLBA data revealed a
complex radio structure and superluminal proper motions with values up
to 4c.  Details are given in Cheung et al. (2007) and we show a
synopsis in fig.~\ref{fig:vlba}.  Extrapolating the observed motions
back in time suggests that the various features originated from the
upstream (quasi-stationary) end of HST-1 in the 2001-2002 time frame.
The obvious conclusion is that the major intensity flare of HST-1 is
closely associated with the 'launching' of super-luminal radio
components from a 'favored location' (the upstream end of HST-1); see
Stawarz et al. (2006) for a discussion of possible reasons as to why HST-1
occurs at this location in the jet.  

\section{The TeV connection}

The H.E.S.S. group reported a higher $\gamma$-ray flux in 2005 than in
adjoining years (Aharonian et al. 2006).  They argued that because of
rapid variability, the likely origin of the TeV emission was the
nucleus (i.e. close to the SMBH).

As discussed in Cheung et al. (2007) the evidence that the TeV excess
emission originated in HST-1 rather than from the nuclear region
includes:

\begin{itemize}

\item{The X-ray emission for HST-1 peaked at the same time as the TeV high state.}

\item{Inverse Compton emission is a mandatory process in any
relativistic plasma.  Both starlight and the synchrotron spectrum peak
near 10$^{14}$ Hz and thus TeV emission is expected from electrons of
Lorentz factor $\approx$10$^6$ (known to exist because of the observed
synchrotron UV to X-ray emission).
The expected intensity is approximately what was observed.}

\item{The IC and synchrotron spectral indices have similar values.}

\item{The photon-photon opacity for TeV photons of the alternate location near the
SMBH is much greater than 1.}

\end{itemize}

\begin{figure}[!ht]
\centerline{\includegraphics[scale=0.4]{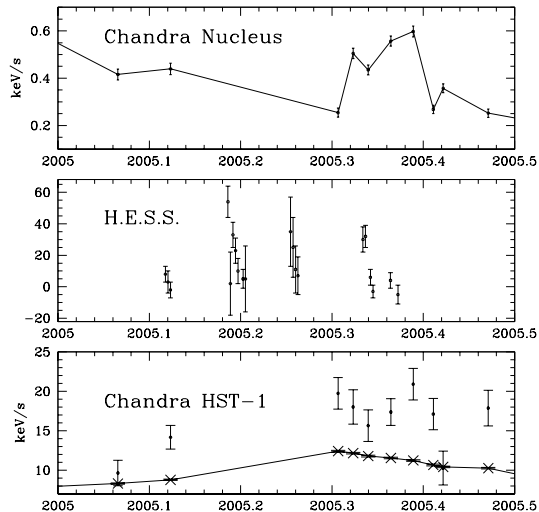}}
\caption{X-ray lightcurves for the nucleus (corrected for 5\%
  contamination from HST-1), HST-1 (top and bottom panels,
  respectively), and the H.E.S.S. TeV data (center panel).
  Readout streak data for HST-1 are included.}\label{fig:3panlc}
\end{figure}

In principle, we might expect to determine the origin of the TeV
emission from a detailed comparison of the light curves of different
components.  The data are ambiguous because of the
contamination suffered by the nucleus and by the gap in
our Chandra coverage during most of the H.E.S.S. fast variability.
In fig.~\ref{fig:3panlc} we show the relevant lightcurves for the
interval of interest in 2005.

\section{Summary}

In general, once we find something
unique like the first pulsar, we then find many more and are able to
study the class of such objects.  In the case of M87, it seems to us
there are two possibilities.


Either the blazar-like behavior of HST-1 is unique and we now
have to search for similar events in other jets (nothing like this has
so far been detected in the Cen A jet, despite many observations over
many years), or M87 is a blazar
and the current paradigm for blazar emission should be modified to
divorce any reliance on the notion that blazar characteristics are
intimately associated with close proximity to the SMBH.


Neither of these options is supported by the failure to detect offsets
between the purported nucleus of blazars and the origin of
superluminal radio components, despite a large accumulation of VLBA
data on quasars and blazars (e.g. the MOJAVE project).

\acknowledgements 

We thank K. Kellermann and M. Lister for comments on VLBA observations
of blazars and acknowledge cooperative involvement with S. Wagner,
D. Horns (HESS); V. Vassiliev, H. Krawczynski, R. Mukerjee (VERITAS);
K. Mannheim and T. Bretz (MAGIC).  We also thank the Chandra
Director's Office for approving a target of opportunity proposal
that allowed us to get several observations of M87 in 2007 February
and March dark times when the TeV systems were observing M87.
These data are included in the lightcurves presented here, but a
detailed analysis is still underway. This work was partially supported
by NASA grants GO5-6118X, GO6-7112X, and GO7-8119X.



%

\end{document}